\documentclass[article,pre,onecolumn]{revtex4}

\usepackage{amsfonts}
\usepackage{amsmath}
\usepackage{amssymb}
\usepackage{graphicx}

\providecommand{\U}[1]{\protect\rule{.1in}{.1in}}

\begin{document}

\title{Standing Electromagnetic Solitons in Degenerate Relativistic Plasmas}
\author{G. Mikaberidze$^{1}$ and V.I. Berezhiani$^{1,2}$}
\affiliation{$^{1}$School of Physics, Free university of Tbilisi, Tbilisi 0131, Georgia}
\affiliation{$^{2}$Andronikashvili Institute of Physics (TSU), Tbilisi 0177, Georgia}
\pacs{52.27.Ny,52.30.Ex,52.35.Mw,42.65.Tg}

\begin{abstract}
The existence of standing high frequency electromagnetic (EM)
solitons in a fully degenerate overdense electron plasma is studied
applying relativistic hydrodynamics and Maxwell equations. The
stable soliton solutions are found in both relativistic and
nonrelativistic degenerate plasmas.
\end{abstract}

\startpage{1}
\endpage{1}
\maketitle

A significant amount of recent publications describe electromagnetic (EM)
waves in relativistic plasmas and majority of them discuss possible roles of
these waves in different astrophysical phenomena. Highly relativistic
plasmas are observed in the cores of white dwarfs \cite{White}, in
magnetosphere of pulsars \cite{Sturok}, in the MeV epoch of the early
Universe \cite{Tajima} and additionally, they probably show up in the
bipolar jets in Active Galactic Nuclei (AGN) \cite{Begelman}.  Plasma can be
relativistic in two following cases: either bulk velocities of fluid cells
should be close to the speed of light, or the kinetic energy of particles
should be greater then their rest energy. In compact objects, such as white
dwarfs and magnetars, the number densities of electrons is believed to be
roughly between $10^{26}$ $cm^{-3}$and $10^{34}$ $cm^{-3}$ \cite{Shapiro},
\cite{Shukla-Rev}. High density plasma can be produced in the laboratory as
well, indeed contemporary petawatt laser systems have the focal intensities $%
I=2\times 10^{22}$ $W/cm^{2}$ \cite{Yanovski}. Moreover, pulses with higher
than $I=10^{26}W/cm^{2}$ intensities are expected to be achieved soon \cite%
{Dunne}.  Superdense plasmas might be formed with densities in the range of $%
10^{23}cm^{-3}$ and $10^{28}cm^{-3}$ \cite{Shukla-Eliasson}, during the
interaction of such EM\ pulses with solid or gaseous targets. Such plasma
will be opaque for conventional laser systems operate at wavelengths $%
\lambda \sim 1\mu m$. The Linac Coherent Light Source (LCLS) is an $X$-ray
free-electron laser produce femtosecond powerful pulses of coherent soft and
hard $X$-rays with wavelengths from $2.2nm$ to $0.06nm$ \cite{X1}.
Exploiting the possibility to focus $X$-ray laser beams on a spot with down
to laser wavelength, the focal intensities $I\simeq 7\times 10^{25}$ $%
W/cm^{2}$ are expected to be reached \cite{X2}. Successful operation of $X$%
-ray free-electron lasers in different centers world wide \cite{Keitel}
opens up new perspectives to study the EM pulse penetration and its
subsequent dynamics in super dense plasma in laboratory conditions.

Highly compressed plasma with an average interparticle distance smaller than
their thermal de Broglie wavelength, can be considered as a degenerate Fermi
gas. When plasma density increases, the more ideal it becomes and the
interactions of its particles can be neglected \cite{Landau}.

EM solitons in classical relativistic plasma is being studied intensively
\cite{Chveni}, but existence and stability of solitary solutions in
degenerate quantum relativistic plasma are investigated mostly for low
frequencies (see \cite{Khan} and references therein).  The publication goal
is to consider existence of a standing, high frequency EM soliton in the
relativistic degenerate electron plasma. Importance of the standing soliton
solutions for overall dynamics of EM pulses is established theoretically
\cite{Bul1} - \cite{Saxena} as well as experimentally for classical
relativistic plasma \cite{Bor}. These publications state, that during
interaction of a circularly polarized strong laser pulse and a plasma, part
of the laser energy is trapped in non-propagating soliton-like pulses.
Similar dynamics is expected in the case of strong EM pulse interaction with
degenerate electron plasma.

Plasma can be considered cold, if the thermal energy of its electrons is
negligible compared to their Fermi energy. In this case temperature can be
assumed to be zero, even though it is of the order of $10^{9}K$ \cite{Russo}%
. For the Fermi energy of electrons we have $\epsilon _{F}=m_{e}c^{2}\left[
\left( 1+R^{2}\right) ^{1/2}-1\right] $, where $R=p_{F}/m_{e}c$ \ and  $p_{F}
$ is the Fermi momentum. The latter is related to the proper density of
electrons $n$ by the following equation $p_{F}=m_{e}c\left( n/n_{c}\right)
^{1/3}$, where $n_{c}=5.9\times 10^{29}cm^{-3}$ is the normalizing critical
number-density \cite{Akbari}. Therefore, when  $n\geq n_{c}$ , electrons
move with relativistic momentum inside plasma unit cells and the plasma can
be told as relativistic.

Our investigation is based on the Maxwell equations and fluid model of
relativistic electron plasma. The ions are considered to form a stationary
neutralizing background. We begin with the manifestly covariant form of the
fluid equations for electrons

\begin{equation}
\frac{\partial T^{\alpha \beta }}{\partial x^{\beta }}=-eF^{\alpha \beta
}nU_{\beta }  \label{B1}
\end{equation}%
here $\partial _{\alpha }=\partial /\partial x^{\alpha }=\left(
c^{-1}\partial /\partial t,\mathbf{\nabla }\right) $; the Greek indexes take
values from $0$ to $3$; $T^{\alpha \beta }$ is the energy-momentum tensor
describing the plasma electrons with charge $-e$, mass $m_{e}$ and the
proper density $n$; the metric tensor is $g^{\alpha \beta }=diag\left(
1,-1,-1,-1\right) $; $U^{\alpha }=\left( \gamma ,\gamma \mathbf{V/}c\right) $
denotes the local four velocity, here $\gamma =\left( 1-V^{2}/c^{2}\right)
^{-1/2}$; $\left( U^{\alpha }U_{\alpha }=1\right) $. This equation implies
the conservation of energy and momentum. The change of momentum through the
collisions is neglected.

We assume, that the total number of electrons is conserved, thus the
following continuity equation is held

\begin{equation}
\frac{\partial nU^{\alpha}}{\partial x^{\alpha}}=0   \label{B2}
\end{equation}

The EM field can be expressed through a tensor $F^{\alpha\beta}=[\mathbf{E},%
\mathbf{B}]$. The Maxwell equations in these notations are $%
\partial_{\beta}F^{\alpha\beta}=-(4\pi/c)J^{\alpha}$ and \qquad$%
\epsilon^{\alpha\beta\gamma\delta}\partial_{\beta}F_{\gamma\delta}=0$. Here $%
J^{\alpha}=(c\rho,\mathbf{J})$, where $\mathbf{J}$ is the current density
and $\rho$ is the total charge density of the plasma.

We use the energy momentum tensor of ideal isotropic fluid $T^{\alpha \beta
}=wU^{\alpha }U^{\beta }-g^{\alpha \beta }P$, where $w=E+P$ is the enthalpy
per unit volume, $P$ is the pressure and $E$ is density of the rest frame
internal energy.   If $nT/P<<1$, plasma can be treated as completely
degenerate Fermi gas and the following equations are satisfied \cite{Chandra}

\begin{equation}
P=\frac{m_{e}^{4}c^{5}}{3\pi^{2}\hbar^{3}}f\left( R\right)   \label{B3}
\end{equation}

\begin{equation}
E=\frac{m_{e}^{4}c^{5}}{3\pi^{2}\hbar^{3}}\left[ R^{3}\left( 1+R^{2}\right)
^{1/2}-f\left( R\right) \right]   \label{B4}
\end{equation}
where

\begin{equation}
8f\left( R\right) =3\sinh^{-1}R+R\left( 1+R^{2}\right)
^{1/2}\left(2R^{2}-3\right)   \label{B5}
\end{equation}

The equation of state for the degenerate gas is $P\propto n^\Gamma$, with $%
\Gamma=5/3$ for nonrelativistic case $\left(R<<1\right)$ and $\Gamma=4/3$
for ultrarelativistic case $\left(R>>1\right)$.

The model of plasma described above implies that the electron distribution
function remains locally Juttner-Fermian. In case of zero temperature this
results in thermodynamical quantities, depending only on density $E(n),P(n)$
and $w(n)$. Of course, these quantities are functions of $x^{\alpha }$
through the relation $n=N/\gamma $, here $N$ is the electron density in the
laboratory reference frame. The considered system is isentropic
(furthermore, as temperature approaches zero, entropy tends to zero too).
Hence, the following thermodynamic equality is held $d\left( w/n\right) =dP/n
$ and taking into account this thermodynamic equality and making some
standard manipulations (e.g. \cite{Ohashi}), Eq. (\ref{B1}) can be
represented in the form of the following system

\begin{equation}
\frac{\partial}{\partial t}\left( G\mathbf{p}\right) +m_{e}c^{2}\mathbf{%
\nabla}\left( G\gamma\right) =-e\mathbf{E+}\left[ \mathbf{V\times \Omega}%
\right]   \label{B6}
\end{equation}

\begin{equation}
\frac{\partial }{\partial t}\mathbf{\Omega }=\mathbf{\nabla \times }\left[
\mathbf{V\times \Omega }\right]   \label{B7}
\end{equation}%
here, for the generalized vorticity we have $\mathbf{\Omega =-}\left(
e/c\right) \mathbf{B+\nabla \times }\left( G\mathbf{p}\right) $ , where $%
\mathbf{p}=\gamma m_{e}\mathbf{V}$ denotes the hydrodynamic momentum; $G=G(n)
$ can be called the density dependent "effective mass" factor of electrons $%
G=w/m_{e}nc^{2}=\left( 1+R^{2}\right) ^{1/2}$. Now dynamics of the
degenerate plasma can be completely described by Eqs. (\ref{B6})-(\ref{B7})
together with Continuity and Maxwell equations. In other words, the set of
equations is complete. The analogous set of equations is derived in Ref.\cite%
{Ohashi} for classical relativistic plasma obeying Maxwell-Juttner
statistics, where $G$ is a function of temperature $G=G(T)$. In contrast,
for degenerate plasma $w/m_{e}nc^{2}=\left( 1+R^{2}\right) ^{1/2}$ and as a
result the effective mass factor of electrons depends just on their proper
density. The corresponding relation $G=\left[ 1+\left( n/n_{c}\right) ^{2/3}%
\right] ^{1/2}$ holds for any ratio $n/n_{c}$, thus for any strength of
relativity \cite{BerScr}-\cite{Beltrami}.

We make use of the expressions for fields $\mathbf{B}=\mathbf{\nabla \times A%
}$ and $\mathbf{E=-(}1/c\mathbf{)\partial A/\partial }t-\mathbf{\nabla }%
\varphi $ where $\mathbf{A}$ and $\varphi $ are vector and scalar potentials
respectively. Applying the Coulomb gauge condition $\mathbf{\nabla \cdot A}=0
$, the Maxwell equations take the following form:

\begin{equation}
\frac{\partial^{2}\mathbf{A}}{\partial t^{2}}-c^{2}\Delta\mathbf{A+}c\frac{%
\partial}{\partial t}\left( \mathbf{\nabla}\varphi\right) -4\pi c\mathbf{J}%
=0   \label{B8}
\end{equation}
and

\begin{equation}
\Delta \varphi =-4\pi \rho   \label{B9}
\end{equation}%
where $\mathbf{J}=-e\gamma n\mathbf{V,}$ is the current density and $\rho
=e(n_{0}-\gamma n)$ is the charge density, with $n_{0}$ denoting electron
(ion) equilibrium density. We use equations (\ref{B6}) and (\ref{B7}) to
describe wave dynamics in unmagnetized plasma. Eq. (\ref{B7}) makes clear
that if generalized vorticity $\mathbf{\Omega }$ was zero everywhere once,
it will stay zero always. Therefore Eq. (\ref{B6}) will reduce to

\begin{equation}
\frac{\partial}{\partial t}\left( G\mathbf{p-}\frac{e}{c}\mathbf{A}\right) +%
\mathbf{\nabla}\left( m_{e}c^{2}G\gamma-e\varphi\right) =0   \label{B10}
\end{equation}

Our goal is to find one dimensional localized solutions for equations (\ref%
{B8})-(\ref{B10}). Let us assume, that every variable depends on nothing but
coordinate $z$ and time $t$. As transverse component of gradient is zero,
Eq. (\ref{B10}) easily gives $\mathbf{p}_{\perp }=e\mathbf{A}_{\perp }/(cG)$%
. Integration constant is zero, because $\mathbf{p}$ should be zero at
infinity, where fields vanish. Coulomb gauge condition requires $A_{z}=0$,
thus the longitudinal motion of the plasma is driven just by the
"ponderomotive" pressure $\left( \sim \mathbf{p}_{\perp }^{2}\right) $
acting via the relativistic $\gamma $ factor in Eq. (\ref{B10}) ($\gamma =%
\left[ 1+\left( \mathbf{p}_{\perp }^{2}+p_{z}^{2}\right) /m_{e}^{2}c^{2}%
\right] ^{1/2}$). The EM pressure forces electrons to move in $z$ direction,
the plasma density changes and charge separation occurs. Longitudinal motion
of the plasma is described by the following set of equations:

\begin{equation}
\frac{\partial}{\partial t}Gp_{z}+\frac{\partial}{\partial z}\left(
m_{e}c^{2}G\gamma-e\varphi\right) =0   \label{B11}
\end{equation}
while the continuity (\ref{B2}) and Poisson's equations (\ref{B9}) become

\begin{equation}
\frac{\partial }{\partial t}\gamma n+\frac{\partial }{\partial z}\left(
n\gamma V_{z}\right) =0  \label{B12}
\end{equation}%
\begin{equation}
\frac{\partial ^{2}\varphi }{\partial z^{2}}=4\pi e(n\gamma -n_{0})
\label{B-12-1}
\end{equation}%
The transverse component of the current density is $\mathbf{J}_{_{\perp }}%
\mathbf{=}\left( ne^{2}/cG\right) \mathbf{A}_{\perp }$ and substituting it
into Eq. (\ref{B8}), we get

\begin{equation}
\frac{\partial^{2}\mathbf{A}_{\perp}}{\partial t^{2}}-c^{2}\frac{\partial
^{2}\mathbf{A}_{\perp}}{\partial z^{2}}\mathbf{+}\Omega_{e}^{2}\left( \frac{n%
}{n_{0}}\frac{G_{0}}{G}\right) \mathbf{A}_{\perp}=0   \label{B13}
\end{equation}
where $n_{0}$ is electron (ion) equilibrium density and $\Omega_{e}=\left(4%
\pi e^{2}n_{0}/m_{e}^{\ast}\right)^{1/2}$ is the Langmiur frequency of the
electron plasma. In this expression $m_{e}^{\ast}$ denotes effective mass of
electron $m_{e}^{\ast}=m_{e}G_{0}$, where $G_{0}=\left[ 1+R_{0}^{2}\right]%
^{1/2}$ and $R_{0}=\left(n_{0}/n_{c}\right)^{1/3}$.

We are searching for solitary stationary solutions of Eqs. (\ref{B11})-(\ref%
{B13}). Assuming electromagnetic wave is circularly polarized, the vector
potential can be rewritten as follows $\mathbf{A}_{\perp}=\left[%
A(z)\cos(\omega t),A(z)\sin(\omega t)\right]$. Here the amplitude $A(z)$ is
a real valued function depending only on coordinate $z$ and nothing else. $%
\omega$ denotes the frequency. It is now convenient to introduce the
following dimensionless quantities: $\Psi=e\varphi/\left(m_{e}c^{2}G_{0}%
\right)$, $a=eA/\left(c^{2}m_{e}G_{0}\right)$, $n=n/n_{0}$, $t=\Omega_{e}t$,
$z=\left(\Omega_{e}/c\right)z$. In our case $p_{z}=0$ and integrating Eq. (%
\ref{B11}) we get the relation

\begin{equation}
\Gamma\gamma=\Psi+1   \label{B14}
\end{equation}
where

\begin{equation}
\Gamma=\frac{G}{G_{0}}=\sqrt{\frac{1+R_{0}^{2}n^{2/3}}{1+R_{0}^{2}}}
\label{B15}
\end{equation}
and $\gamma=\left(1+a^{2}/\Gamma^{2}\right)^{1/2}$. Using Eqs. (\ref{B14})-(%
\ref{B15}) we obtain the following relationships:

$\bigskip$%
\begin{equation}
n=\frac{1}{R_{0}^{3}}\left[(1+R_{0}^{2})\left[\left(1+\Psi\right)^{2}-a^{2}%
\right]-1\right]^{3/2}  \label{B16}
\end{equation}
\begin{equation}
\gamma=\frac{\left(1+\Psi\right)}{\sqrt{\left(1+\Psi\right)^{2}-a^{2}}}
\label{B17}
\end{equation}
while for the electron density in laboratory frame $\mathcal{N}%
\left(=n\gamma\right)$ we have

\bigskip

\begin{equation}
\mathcal{N}=\frac{\left(1+\Psi\right)}{\epsilon^{3/2}\left[%
\left(1+\Psi\right)^{2}-a^{2}\right]^{1/2}}\left[\left(1+\Psi%
\right)^{2}-(1+a^{2}-\epsilon)\right]^{3/2}  \label{B18}
\end{equation}
where $\epsilon=R_{0}^{2}/\left(1+R_{0}^{2}\right)$. \bigskip

The Eqs. (\ref{B-12-1}) and (\ref{B13}) now can be reduced to the following
set of ordinary differential equations:

\begin{equation}
\frac{d^{2}a}{dz^{2}}-(1-\omega^{2})a+\left(1-\frac{\mathcal{N}%
\left(a,\Psi,\epsilon\right)}{1+\Psi}\right)a=0   \label{B19}
\end{equation}
\begin{equation}
\frac{d^{2}\Psi}{dz^{2}}+1-\mathcal{N}\left(a,\Psi,\epsilon\right)=0
\label{B20}
\end{equation}

The coupled system of nonlinear equations (\ref{B18})-(\ref{B20}), where $%
\mathcal{N}\left(a,\Psi,\epsilon\right)$ is defined by Eq. (\ref{B18})
describes the structure of circularly polarized transverse and longitudinal
localized fields in degenerate plasma for arbitrary values of $R_{0}$ $%
\left(0\leq\epsilon\leq1\right)$. The general solution of Eqs. (\ref{B18})-(%
\ref{B20}) cannot be expressed in terms of elementary functions, it requires
numerical methods. However, these equations can be integrated analytically
for certain limiting cases. Taking into account that for $\left\vert
z\right\vert\rightarrow\infty$ the fields vanish $a\rightarrow0,
\Psi\rightarrow0$ and $\mathcal{N}\rightarrow1$ we conclude that the
standing solitary solutions may exist in an overdense plasma when the
frequency of the soliton is less than the electron plasma frequency, $%
\omega<1$ ($\omega<\Omega_{e}$ - in dimensional units). For any fixed plasma
density (i.e. fixed $\epsilon$) the frequency $\omega$ is the only parameter
in Eqs. (\ref{B18})-(\ref{B20}) that defines the characteristics of the
solitary solutions. For slightly overdense plasma $\omega\leq1$ the fields
are small $a\ll1$, $\Psi\ll1$ while the characteristic width of the
structure $L\sim(1-\omega^{2})^{-1/2}>>1$. Applying the quasi-neutrality
condition of the plasma ($\mathcal{N\approx}1$) from Eqs. (\ref{B18}) and (%
\ref{B20}) we get $\Psi\simeq\left(a^{2}/2\right)\left(1-\epsilon/3\right)$
and now Eq. (\ref{B19}) reduces to the simple equation

\begin{equation}
\frac{d^{2}a}{dz^{2}}-(1-\omega^{2})a+\frac{1}{2}\left(1-\epsilon/3\right)
a^{3}=0  \label{B21}
\end{equation}
with the well know soliton solution

\begin{equation}
a=2\sqrt{\frac{1-\omega^{2}}{1-\epsilon/3}}Sech\left[ z\sqrt{1-\omega^{2}}%
\right]   \label{B22}
\end{equation}

Analytical solutions of Eqs. (\ref{B18})-(\ref{B20}) can be obtained even
for an arbitrary amplitude fields provided that plasma density is small $%
n_{0}/n_{c}<<1$ ($R_{0}<<1$), implying that the degenerate electron gas is
not relativistic. At this end we would like to emphasize that our
consideration is valid if the average kinetic energy of electrons ($%
\sim\epsilon_{F}$) is larger than their interaction energy ($\sim
e^{2}n_{0}^{1/3}$). This condition is fulfilled for a sufficiently dense
plasma when $n_{0}>>\left(2m_{e}e^{2}/\left(3\pi^{2}\right)^{2/3}\hbar^{2}%
\right)^{3}=6.3\times10^{22}cm^{-3}$. For the densities $n_{0}$ in the range
from $10^{24}cm^{-3}$ to $10^{28}cm^{-3}$ the corresponding parameter $%
\epsilon$ is in the range from $1.4\times 10^{-4}$ to $6.2\times10^{-2}$.
Since $\epsilon$ is rather small for the nonrelativistic densities, we can
safely assume that $\epsilon\rightarrow0$ in Eqs. (\ref{B18})-(\ref{B19})
and then obtain the following relations:

\begin{equation}
\Psi=1-\sqrt{1+a^{2}}  \label{B23}
\end{equation}
\begin{equation}
\mathcal{N}=1+\frac{d^{2}}{dz^{2}}\sqrt{1+a^{2}}  \label{B24}
\end{equation}
the wave equation (\ref{B20}) now reduces to

\begin{equation}
\frac{d^{2}a}{dz^{2}}+\omega^{2}a-\frac{a}{\sqrt{1+a^{2}}}\left(1+\frac{d^{2}%
}{dz^{2}}\sqrt{1+a^{2}}\right)=0  \label{B25}
\end{equation}

Similar to (\ref{B23})- (\ref{B25}) system equations have been derived
solved in Refs. \cite{Kurki}, \cite{Bulanov} where existence and stability
of standing EM pulses in cold but classical electron plasma were studied.
The soliton solution of Eq. (\ref{B25}) is found to be

\bigskip%
\begin{equation}
a=\frac{2\sqrt{1-\omega^{2}}\cosh\left( z\sqrt{1-\omega^{2}}\right)}{%
\cosh^{2}\left(z\sqrt{1-\omega^{2}}\right)+\omega^{2}-1}  \label{B26}
\end{equation}

The amplitude of this single hump soliton $a_{m}(=a(0))$ depends on the
soliton frequency by the relation $a_{m}=2\sqrt{1-\omega^{2}}/\omega^{4}$.
The electrons are depleted from the region of pulse localization. The
minimal density at the center of soliton is $\mathcal{N}(z=0)=1-4\left(1-%
\omega^{2}\right)^{2}/\omega^{4}$\cite{Bulanov}. Note that for $%
\omega\rightarrow1$ the amplitude of soliton decreases and the solution (\ref%
{B26}) coincides with Eq. (\ref{B22}) where $\epsilon<<1$. With a decrease
of $\omega$ the soliton amplitude increases while the corresponding density
well deepens. For $\omega=\omega_{c}=\sqrt{2/3}=\allowbreak0.816$ the
electron cavitation takes place $\mathcal{N}(z=0)=0$ while the amplitude of
the soliton attains its maximal value $a_{m}=\sqrt{3}$. For $a_{m}>\sqrt{3}$
the solution contains a region where the electron density is negative which
implies that wave breaking takes place. Thus, in the nonrelativistic
degenerate electron plasma the standing solitons exist provided $%
\omega_{c}\leq\omega\leq1$. Similar conclusion has been made in Ref.\cite%
{Eliason} by numerical analysis of the coupled set of nonlinear
Schrodinger-Poisson equations, where the Bohm potential was introduced in
the system. In the fluid approach the Bohm potential is related to the
"quantum force" due to electron tunneling. Our estimations show that this
force is small and even for $\omega=\omega_{c}$ it does not change the
solution (\ref{B26}) qualitatively.

Now we consider the case of ultrarelativistic degenerate plasma $R_{0}>>1$ ($%
\epsilon\rightarrow1$). The system (\ref{B18})-(\ref{B20}) reduces to the
following set of equations

\begin{equation}
\frac{d^{2}a}{dz^{2}}+\omega^{2}a+a\left( a^{2}-\left( 1+\Psi\right)
^{2}\right) =0   \label{B27}
\end{equation}
and

\begin{figure}
  \centering {\includegraphics[width=7cm]{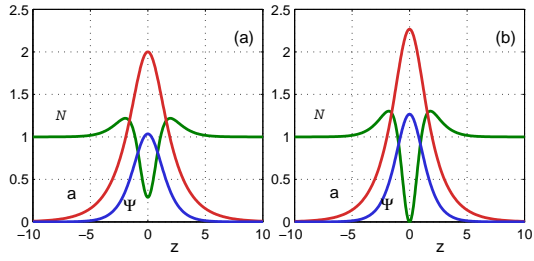}}
  \caption{The profiles of EM field potentials $a$, $\Psi$ and
the electron density $\mathcal{N}$ in case the ultrarelativistic plasma $%
\epsilon=1$ for the different soliton frequencies (a) $\omega=0.8$ and (b) $%
\omega=0.7683$.}\label{fig1}
\end{figure}

\begin{figure}
  \centering {\includegraphics[width=7cm]{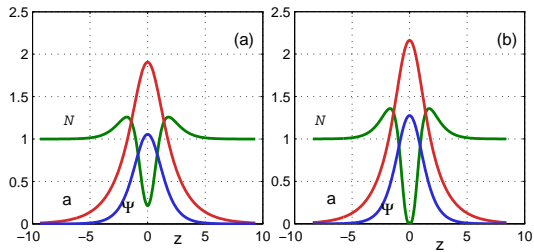}}
  \caption{The profiles of EM field potentials and the electron
density in case the relativistic plasma with $\epsilon=0.5$ for the
different soliton frequencies (a) $\omega=0.8$ and (b)
$\omega=0.7694$.}\label{fig2}
\end{figure}
\begin{equation}
\frac{d^{2}\Psi}{dz^{2}}+1+\left( 1+\Psi\right) \left( a^{2}-\left(
1+\Psi\right) ^{2}\right) =0   \label{B28}
\end{equation}

This system of equations has a first integral (a Hamiltonian)

\begin{equation}
\left[ \frac{d\Psi}{dz}\right] ^{2}-\left[ \frac{da}{dz}\right]
^{2}-\omega^{2}a^{2}+2\Psi-\frac{1}{2}\left[ a^{2}-\left( 1+\Psi\right) ^{2}%
\right] ^{2}+\frac{1}{2}=0   \label{B29}
\end{equation}
where the zero boundary conditions for the fields $\Psi$, $a$ and their
derivatives at $\left\vert z\right\vert\rightarrow\infty$ have been used.

Fig.1 shows the numerical solution of the system (\ref{B18})-(\ref{B20}) for
different values of the soliton frequency $\omega$. One can see that the
structure of the solitary solutions in ultrarelativistic plasma is similar
to one, obtained in the nonrelativistic case. The solution has a continuous
spectrum with $\omega_{c}\leq\omega\leq1$ where $\omega_{c}=0.7683$, and it
is composed of a single maximum of field potentials ($a,\Psi$) and depleted
electron density ($\mathcal{N}$) region at the center of structure. For $%
\omega\rightarrow1$ the numerical solution coincides with Eq. (\ref{B22})
(using $\epsilon=1$). For $\omega\simeq\omega_{c}$, amplitude of the soliton
becomes relativistically strong $a_{m}\simeq2.267$, while at the center of
the soliton $\mathcal{N}\simeq0$, i.e. the electron cavitation takes place.

Numerical analysis demonstrates that equations (\ref{B19})-(\ref{B20}) have
soliton-like solutions for any finite value of $R_{0}$. As an example in
Fig.2 we plot the numerical solutions of the system for $R_{0}=1$ ($%
\epsilon=0.5$). The soliton exists for $\omega_{c}\leq\omega\leq1$, where $%
\omega_{c}=0.7694$ and at this frequency $a_{m}\simeq2.162$.

To verify the stability of obtained solutions, we performed 1D simulations
applying the numerical code developed in Ref.\cite{Berezhiani}. The
simulations demonstrate exceptional stability of the solution for entire
rage of soliton existence described above.

To summarize, in this paper we considered circularly polarized high
frequency EM solitons in degenerate electron plasma. We used Maxwell and
relativistic fluid equations to demonstrate possibility of stable existence
of solitons in overdense plasma ($\omega \leq \Omega _{e}$). Soliton exists
for entire range of physically allowed electron densities, presumably for $%
n_{0}=10^{24}cm^{-3}$ and higher, i.e. in both relativistic and
nonrelativistic degenerate plasmas. Intensity of the solitons can be small
for $\omega \rightarrow \Omega _{e}$ and becomes relativistically strong ($%
a_{m}>1$) for $\omega \rightarrow \omega _{c}$. Appears to be, that
cavitation of plasma takes place in both, relativistic and nonrelativistic
degenerate plasmas. The model described above can be straightforwardly
generalized for underdense plasma. This is not done in the scope of this
paper deliberately.

The present results could have rather interesting implications to describe
and understand $X$-ray pulses emerging from compact astrophysical objects.
It is also of particular interest to describe interaction of intense laser
pulses and dense degenerate plasma, as such experiments are becoming
feasible with the new generation of intense lasers.

\bigskip

\bigskip

\bigskip

\newpage

%{\LARGE Figure Captions}

%\medskip

%FIG.1. (Color online) The profiles of EM field potentials $a$, $\Psi$ and
%the electron density $\mathcal{N}$ in case the ultrarelativistic plasma $%
%\epsilon=1$ for the different soliton frequencies (a) $\omega=0.8$ and (b) $%
%\omega=0.7683$.

%\bigskip

%FIG.2. (Color online) The profiles of EM field potentials and the electron
%density in case the relativistic plasma with $\epsilon=0.5$ for the
%different soliton frequencies (a) $\omega=0.8$ and (b) $\omega=0.7694$

\end{document}